\documentclass[a4paper]{article}
\usepackage{verbatim}
\usepackage[left]{lineno}
\usepackage{amsfonts}
\usepackage{authblk}
\newtheorem{theorem}{Theorem}
\newtheorem{lemma}[theorem]{Lemma}

\newtheorem{definition}{Definition}
\title{Optimal Las Vegas reduction from one-way set reconciliation 
to error correction}
\author{Djamal Belazzougui\thanks{This work was done while the 
author was working at the University of Helsinki.}} 
\affil{DTISI, CERIST Research center, 05 rue des trois freres Aissou, Benaknoun, Algiers, Algeria}

\begin{document}
\maketitle
\begin{abstract}
Suppose we have two players $A$ and $C$, where player  $A$
has a string $s[0..u-1]$ and player $C$ has a string $t[0..u-1]$ 
and none of the two players knows the other's string. 
Assume that $s$ and $t$ are both over an integer alphabet
$[\sigma]$, where the first string contains $n$ non-zero entries. 
We would wish to answer to the following basic question. Assuming that $s$ 
and $t$ differ in at most $k$ positions, how many bits does player $A$ need 
to send to player $C$ so that he can recover $s$ with certainty? 
Further, how much time 
does player $A$ need to spend to compute the sent bits and how much time does player $C$ need 
to recover the string $s$? 

This problem has a certain number of applications, 
for example in databases, where each of the two parties possesses 
a set of $n$ key-value pairs, where keys are from the universe $[u]$ and 
values are from $[\sigma]$ and usually $n\ll u$. 
In this paper, we show a time and message-size optimal Las Vegas reduction from this problem to the 
problem of systematic error correction of $k$ errors for strings of length $\Theta(n)$ over an 
alphabet of size $2^{\Theta(\log\sigma+\log (u/n))}$.  
The additional running time incurred by the reduction is linear randomized for player $A$ and linear deterministic for 
player $B$, but the correction works with certainty. 
When using the popular Reed-Solomon codes, the reduction gives a protocol
that transmits $O(k(\log u+\log\sigma))$ bits and runs in time $O(n\cdot\mathrm{polylog}(n)(\log u+\log\sigma))$ 
for all values of $k$. The time is randomized for player $A$ (encoding time) and 
deterministic for player $C$ (decoding time). The space is optimal whenever $k\leq (u\sigma)^{1-\Omega(1)}$.

\end{abstract}
\section{Introduction}
Suppose we have two strings $s$ and $t$ of equal length $u$ over an 
integer alphabet $[\sigma]$, where the number of non-zeros in the first string equals $n$
and the Hamming distance between the two strings is at most $k$
(the number of positions $i$ such that $s[i]\neq t[i]$)

Moreover suppose that we have two players $A$ and $C$, 
where player $A$ knows 
$s$, but does not know $t$ and player $C$ 
knows $t$ but does not known $s$. That is, the only information
a player has about the other's string is the upper 
bound $k$ on the Hamming distance between the two strings. 

Then we would like to design a one-way protocol in which 
player $A$ computes some data on his string $s$ and sends it to player $C$, 
so that the latter can recover the string $s$. How good can such a 
protocol be, in terms of the size of the sent data and in terms of 
time complexity? 

This problem can be considered as a variant of the \emph{document exchange} problem, 
which is defined as follows. 
Player $A$ has a document $s$ and player $C$ has a document 
$t$ and the only knowledge that a player has 
about the other player's document is 
an upper bound on some distance between 
the two documents (the distance can be edit distance,
Hamming distance or some other distance). 
Then the two players must exchange some data 
so that each of the two can recover 
the document that is held by the other. 
Other variants of the problem have already been studied
in the literature, most notably document exchange under the edit 
distance (see paper~\cite{Jo12} and references 
therein).


The problem is also known in the literature 
as the \emph{set reconciliation} problem~\cite{MinskyTZ03,goodrich2011invertible}.
In the database context, the problem can be considered as that 
of synchronizing two sets of $n$ key-value pairs, 
where the keys are from universe $[u]$ and values are 
from the set $[\sigma]$. 

The problem can be 
solved with optimal message size (up to constant factor) using 
traditional error correcting codes, like Reed-Solomon codes~\footnote{
An optimal message-size requires a code with $\Theta(k\log(u\sigma/k))$ 
bits of redundancy. Since Reed-Solomon codes have 
$O(k\log(u\sigma))$ bits of redundancy they will lead to optimal message size $O(k\log(u\sigma/k))$ bits only if $k\leq (u\sigma)^{1-\epsilon}$ for some constant $\epsilon>0$. 
Notice however, there provably exist
less time efficient (exponential time) codes that achieve the required redundancy
~\cite{gilbert1952comparison,varshamov1957estimate} for all values of $k$.}. 
However, those solutions would require time $\Omega(u)$. 
In many cases $n\ll u$ and time $\Omega(u)$ is 
unacceptable (for example in database context, 
we could have $u=2^{64}$ or $u=2^{128}$ with $n$ ranging
from $2^{10}$ to $2^{30}$). 

It is thus much more preferable to have a time that depends on $n$
(the number of non-zero elements) rather than $u$. 
The best one could wish for is a time linear in $n$, 
since even reading or writing a representation 
of the strings $s$ or $t$ (even in 
compressed form) requires $\Theta(n)$ time in the worst-case. 

There already exist solutions to the problem with 
time bounds that depend only on $n$ and $k$ (possibly 
with logarithmic dependence 
on $u$). We review some of those solutions. In all cases  
an alphabet size $\sigma=2$ is assumed, but the solutions can easily be 
extended to support an arbitrary alphabet by replacing 
$u$ by $u\sigma$~\footnote{\label{big_to_binary_alpha}That is, given $s$ (and $t$) one can construct a string 
of length $u\sigma$, by expanding every element $s[i]$ (resp. $t[i]$) into a unitary binary vector 
of length $\sigma$ that contains a single $1$ at position $s[i]$ (resp. $t[i]$).}. 

The deterministic solution proposed in~\cite{MinskyTZ03} uses polynomials 
over finite fields. The protocol
transmits $O(k\log u)$~\footnote{In this paper we define $\log x=\lceil\log_2(\mathrm{max}(x,2))\rceil$.} bits and has time complexity $O(nk+k^3\log u)$. 
The term $nk$ is needed for evaluating a polynomial at $k$ given points. We note that the term can be 
improved to $O(n\cdot\mathrm{polylog}(u))$ by using faster (and more sophisticated) 
operations on polynomials (e.g. using the discrete Fourier transform~\cite{mateer2008fast}). 
The term $k^3\log u$ comes from the rational function interpolation 
and the factoring of two polynomials of degree $k$. The best algorithm currently known
for factoring a polynomial of degree $k$ runs in time $\Omega(k^{3/2})$~\cite{kedlaya2011fast}. 
Thus the total time will be $n^{1+\Omega(1)}$ when $k$ is at least $n^{2/3+\Omega(1)}$. 

A randomized Monte-Carlo solution achieving $O(n\log u)$ running time 
and $O(k\log u)$ transmitted bits was achieved in~\cite{porat2007improved}. That solution only guarantees 
a successful reconciliation with high probability. 
Finally approaches based on Invertible Bloom filters
~\cite{eppstein2007space,goodrich2011invertible,eppstein2011s,mitzenmacher2012biff} achieve $O(n)$ time
with $O(k\log u)$ bits, but again working with only probabilistic guarantees~\footnote{The 
analysis of those solutions is not fully rigorous though, as the solutions 
do not specify the used hash functions.}. 

The main result of this paper is to show a  
time and message-size optimal Las-Vegas reduction from our 
problem to the problem of systematic error correction of $k$ 
errors for strings of length $\Theta(n)$ over an 
alphabet of size $2^{\Theta(\log\sigma+\log (u/n))}$. 

The high level idea is to use the two-level 
hashing scheme of~\cite{FredmanKS84} in which the 
second level is made deterministic using 
the deterministic hashing results from~\cite{raman1996priority} 
and~\cite{fredman1993surpassing}. 

This first step takes $O(n)$ expected time in total. 
Then player $A$ applies a systematic error correction 
on the resulting data structure built on his string $s$ 
and sends the redundancy to player $C$. 
Finally player $C$ uses the redundancy he receives from player $A$ 
along with the knowledge of his own string $t$ to recover 
in (deterministic) linear time 
the two level hashing data structure that was built by player $A$
~\footnote{For this step to run in deterministic 
time, it it crucial that the second level of the hashing is deterministic.}. 
The latter allows player $C$ to recover the whole string $s$.

As a corollary of our main result, we show the first (to the 
best of our knowledge) Las-Vegas solution with 
$O(k(\log u+\log\sigma))$ transmitted bits and time $O(n\cdot\mathrm{polylog}(n)(\log u+\log\sigma))$
independent of $k$. The time is randomized for player $A$ and 
worst-case for player $C$.

\section{Preliminaries}

\subsection{Model and Notation}
We assume a word RAM model with word size $w=\Omega(\log u+\log\sigma)$, 
and in which all standard arithmetic and logic operations (including 
multiplication but not division) are assumed to take constant time. 
All the considered alphabets are integer alphabets 
$[\beta]$, where $\beta$ is called the alphabet size 
and $\log\beta$ is called the bitlength of the 
alphabet. 
We use $[\alpha]$ to denote the set $[0..\alpha-1]$. 
We denote the string of player $A$ by $s$ and the 
string of player $C$ by $t$. The length of the first 
string is denoted by $n$ and the maximal Hamming 
distance between the two strings by $k$. 
Throughout it is implicitly assumed that the
strings are given in the following compressed form. 
The string $s$ is represented by 
a list of $n$ pairs $(p,s[p])$ where $s[p]\neq 0$. 
Similarly $t$ is represented by a list of pairs that 
represents all non-zero positions in $t$ along with their 
values. Both lists can be in arbitrary order.

\subsection{Tools}
We now present the main tools that will be used in our 
constructions. 
Given a length $n$ and an error threshold $k$, 
an error correcting code encodes any string $s$ of length $n$ 
into a longer string $s'$ such that $s$ can be recovered from $s'$, 
even if up to $k$ arbitrary positions in $s'$ are changed.  
A code is said to be systematic if the encoded message $s'$ 
includes the original message $s$ verbatim. The sequence of
$|s'|-|s|$ additional symbols that are added to $s$ to form $s'$ 
is called the \emph{redundancy} of the code. 
Given an error correcting code capable of correcting $k$ errors out 
of a string of length $n$ over alphabet of size $\sigma$, we denote the redundancy by 
$r(n,k,\sigma)$, the encoding time by $t_e(n,k,\sigma)$ and 
the decoding time by $t_d(n,k,\sigma)$.

Reed-Solomon codes~\cite{reed1960polynomial} are among the most popular 
codes, achieving optimal redundancy (up to constant factors), when the 
alphabet of the strings is sufficiently large.  

We will use the following result about the encoding and decoding of Reed-Solomon codes: 
\begin{lemma}~\cite{justesen1976complexity,pan1997faster}
\label{lemma:RS_enc_dec}
There exists a systematic error correcting code that encodes 
a string $s$ of length $n$ over an alphabet $[n]$ into a string 
$s'$ of length $n+\Theta(k)$ over an alphabet of size $\Theta(n)$ 
and that allows to correct up to $k$ errors in $s'$. 
Moreover the encoding and decoding can be done in $O(n\cdot \mathrm{polylog} (n))$
\end{lemma}

Some of our results will make use of two special-purpose 
instructions defined as follows: 

\begin{definition}
Given a number $xx_1x_2\ldots x_r\in[2^r]$, 
we define the instruction $\mathrm{MSB}(x)$
as the one that returns the that isolates the most significant set bit. 
That is, given $\mathrm{MSB}(x)=y$ if and only if $x=x_1x_2\ldots x_{y-1}10^{r-y}$
\end{definition}

\begin{definition}
that given an integer 
Given $x=x_1x_2\ldots x_r\in[2^r]$ and mask $M\in[2^r]$, 
We define the instruction $\mathrm{PACK}(x,M)$ as the function 
returning the number $x'=x_{i_1}x_{i_2}\ldots x_{i_t}0^{r-t}$, 
where $i_1<i_2<\ldots <i_t$ are the positions of bits set to $1$ in $M$ .
\end{definition}

We will use the following two results about randomized hashing. 

\begin{lemma}
\label{lemma:2wise_hash_no_prime}~\cite[consequence of Theorem 3]{dietzfelbinger1996universal}
Given two numbers $u$  and $n$ with $n<u$ and 
a set $S\subset [u]$ with $|S|=n$, we can  
find in expected $O(n)$ time two numbers $a,b\in[u\cdot n]$, such that $f(x)\neq f(y)$
for all $x,y\in S$ and $x\neq y$ where $f$ 
is a function from $[u]$ into $[n^2]$ such that
$f(x)=((x\cdot a+b)\;\mathrm{div}\;(u\cdot n^2))\;\mathrm{div}\;u$.
\end{lemma}
The lemma derives from the fact that the 
family of hash functions $f$ parametrized by $a$ and $b$ 
is $2$-wise independent (~\cite[Theorem 3]{dietzfelbinger1996universal}).  

\begin{lemma}
\label{lemma:3wise_hash}

For an arbitrary given prime $P$, a number
$n$ with $n^2<P$ and a set $S\subseteq[P]$ with $|S|=n$
one can in expected time $O(n)$ find three integers $a,b,c\in[P]$ such that
for the function $f(x)=(ax^2+bx+c)\bmod P$ the following holds: 

\begin{enumerate}
\item $(f(x)\bmod n^2)\neq (f(y)\bmod n^2)$ for all pairs $(x,y)\in S^2$ with $x\neq y$. 
\item $\sum_{i=0}^{n-1} |S_i|^3=\alpha\cdot n$, where 
$S_i=\{x\; |\; x\in S , (f(x)\bmod n)=i\}$. 
\end{enumerate}
\end{lemma}
The lemma derives easily from the fact that the 
family of hash functions $f$ parametrized by $a$, $b$ 
and $c$ is $3$-wise independent~\cite{wegman1981new}. That is, 
by $2$-wise independence, the probability 
of not fulfilling first condition is at most $1/2$
and by $3$-wise independence the probability of not fulfilling second condition 
can be made at most $1/4$ by appropriate choice of $\alpha$. Thus, by union 
bound the probability for both conditions to be false is at most $3/4$, 
and hence appropriate values for parameters $a,b,c$ can be be found after an expected 
constant number of trials, each time checking whether the condition are fulfilled 
in time $O(n)$. Thus the expected time to find appropriate values for $a,b,c$ is $O(n)$. 

We will also use the following two (well-known) results about deterministic hashing:

\begin{lemma}
\label{lemma:Ra96}~\cite[consequence of Lemma 6]{raman1996priority}

Given two numbers $r$  and $s$ with $s<r$ and 
a set $S\subset [2^r]$ with $|S|=n\leq 2^{s/2}$, then it is possible 
to find in deterministic $O(n^2r)$ time two numbers $a,b\in[2^r]$, such that $f(x)\neq f(y)$
for all $x,y\in S$ and $x\neq y$ with $f(x)=((x\cdot a+b)\;\mathrm{mod}\;2^r)\;\mathrm{div}\;2^{r-s}$.
\end{lemma}

\begin{lemma}
\label{lemma:Fusion_tree}
Given two integers $r$ and $n$ with $n\leq r\leq w$, 
there exists a family $F$ of hash functions from $[2^r]$ into $[2^{n-1}]$ 
such that a member $f$ of the family can be described using $O(r)$ bits,
 $f(x)$ can be evaluated in constant time and moreover given 
any set $S\subset [2^r]$ of size $n$, one can in deterministic 
$O(n\log n)$ time find a member $f$ of the family such that $f(x)\neq f(y)$ for 
any $x,y\in S$ with $x\neq y$. The construction time assumes the availability 
of an instruction $\mathrm{MSB}$ that can be evaluated in constant time. 

The evaluation time assumes the availability of an instruction $\mathrm{PACK}$ 
that can be evaluated in constant time. 
\end{lemma}
The construction is done as follows. 
The perfect hash function is simply the concatenation 
of (at most) $n-1$ selected bits out of the $r$ bits.
The hash function description consists in a
word $A$ considered as a bitvector that marks (with a $1$) 
the selected bits. The description 
is built as follows. Initially all bits in $A$ are set to zero. 

We then sort the elements of the set $S$ in time $O(n\log n)$. 
We let that order be $x_1,x_2,\ldots,x_n$. We then isolate the most significant set 
bit in $(x_i\;\mathrm{XOR}\;x_{i+1})$ for all $i\in[1,n-1]$ (using the $\mathrm{MSB}$ instruction) 
and OR the result with $A$. The end result is a word $A$ 
that contains $n'\leq n-1$ ones. 
Then evaluating $f(x)$ amounts to computing the word $x'$
which contains in its least $n'$ significant bits the bits in $x$ whose corresponding positions
are set to one in $A$ (using the $\mathrm{PACK}$ instruction). 
The function $f$ will be perfect since every pair of keys will differ in 
at least one of the selected $n'$ positions~\cite{fredman1993surpassing}.

\section{The reduction}

We establish a general reduction from the set reconciliation on strings of length $u$ 
over an alphabet $[\sigma]$, where one of the two strings has $n$ non-zeros 
and the two strings differ in at most $k$ positions
to three instances of systematic error encoding that correct $k$ errors on strings of length $\Theta(n)$ over alphabets of 
bitlength (at most) $\Theta(\log (u/n)+\log\sigma)$. 
The reduction is error-free and uses $O(n)$ randomized time (i.e. the reduction is Las-Vegas)
for the sender and deterministic $O(n)$ time for the reader. 

The main idea of the reduction is to use the framework 
of two-level hashing introduced in~\cite{FredmanKS84} 
and known as $\emph{FKS}$ with a careful choice of hash functions and combining it 
with systematic error correction encoding. 

More precisely the sender will produce a two level hashing 
The first level consists in a global hash function that maps the 
$n$ non-zero positions of its string into an array 
of $n$ buckets. The second level is an array of 
$n$ hash functions where each hash function is specific to a bucket and 
will map the 
$m$ keys of the bucket injectively into the range $[m^2]$. 
By a careful choice of first level and second level hash functions, 
we can ensure that the total construction time is linear and that 
the sum of the range sizes of all buckets will also be linear. 
The second level of the hashing scheme will provide an injective 
mapping from positions into a range of size $n'=\Theta(n)$ and the final step will consist 
in storing in a table of size $n'$ the non-zero positions and their value (character). 
The sender then appliers systematic error correcting codes 
on three tables: a table that stores the bucket hash functions, a table 
that stores the number of keys assigned to each bucker and the 
table of pairs of non-zero positions and their values. 
Finally the sender sends a description of the first level hash 
function and the redundancies built on 
the three tables. 
The receiver will then rebuild a two-level hash function scheme 
on its own keys, but taking care of doing it progressively in three 
steps where at each step the output is corrected using the received 
redundancy. That is at first step, the receiver simply uses the 
global hash function he received. He then corrects the table 
that stores the number of keys assigned to each bucket. 
He then builds the hash functions specific to each bucket, but 
ignoring the buckets that have a different size from the sender. 
After that he corrects the table of bucket hash functions with the received 
redundancies and applies the hash function on its own keys to build 
the table of positions and values and corrects it using the corresponding 
redundancy of the sender. Finally reconstructing the string of the sender 
can be done by scanning the corrected table of positions and values. 
An attentive reader might wonder whether the following simple approach 
would work: build the whole $\emph{FKS}$ scheme on both the sender and 
receiver side and reconcile them using an error correcting code built 
on the concatenation of the FKS components of the sender . 
We argue that this approach will potentially have 
much more $O(k)$ errors. For example, if one uses the different global hash functions, 
then the resulting buckets would be completely different, even if the set of keys 
is the same. Likewise using different bucket hash functions will result in a table 
of pairs and values containing potentially much more than $k$ mismatches. 
Thus synchronizing the components one-by-one is really necessary to ensure that the 
number of errors at next component is upper bounded by $k$.


\subsection{Sender protocol}

The reduction uses the two-level hashing scheme of~\cite{FredmanKS84}.
Player $A$ builds the two-level hashing scheme on its set of $n$ keys
over a universe of size $u$, where the keys are the positions of the non-zeros in $s[0..u-1]$. 

The first level consists in a hash function $g_1$ 
computed using Lemma~\ref{lemma:2wise_hash_no_prime} and 
a hash function $g_2$ computed using Lemma~\ref{lemma:3wise_hash}
(hash function $f$). 
The hash function $g_1$ is computed on the set $S$ mapping $S$ 
injectively into $S'\subset[n^2]$ and then 
$g_2$ is computed on $S'$. 

In order to apply Lemma~\ref{lemma:3wise_hash}, we first find a prime $P$
in $[n^2,2n^2-1]$ using a method described in~\cite{dietzfelbinger1997reliable}. 
This is done in time $o(n)$ by repeatedly selecting 
a random number in $[n^2,2n^2-1]$ and testing whether it is prime or
not using a deterministic primality testing algorithm in~\cite{agrawal2004primes}. 
Since the density of the primes in the interval $[n^2,2n^2-1]$ is $\Theta(1/\log n^2)=\Theta(1/\log n)$ 
and the primality testing algorithm runs in time $O(\mathrm{polylog}(n))$, 
we conclude that the time to find a prime is $O(\mathrm{polylog}(n))$. 

We let $f$ be such that $f(x)=g_2(g_1(x))$. 
The hash function description occupies $O(\log u+\log P)=O(\log u)$ bits. 
We additionally define the functions $f_1$ and $f_2$ as follows. 
We let $f_1(x)=(f(x)\bmod n)$ and $f_2(x)=(f(x)\bmod n^2)$. 
A key $x$ will be mapped to bucket number $i$ if and only if $f_1(x)=i$. 
Denote by $b_i$ the number of keys mapped to bucket $i$. 
Then by Lemma~\ref{lemma:3wise_hash} we 
will have the guarantee that $\sum_{i=0}^{n-1}(b_i)^3=O(n)$. 
The lemma also guarantees that the function $f_2$ is injective. 
We could have used the lemma directly on the set $S$, and we would have 
obtained the same result. The only reason we did not do so, is to 
avoid the $O(\mathrm{polylog}(u))$ cost associated with the search for 
a prime number in $[u,2u-1]$. 
The second level of the scheme is implemented over each bucket. Each bucket will use 
a local hash function $h_i$ determined using lemmata~\ref{lemma:Fusion_tree}
and~\ref{lemma:Ra96}. The
hash function $h_i$ maps the $b_i$ keys to $4(b_i)^2$ cells
(the factor $4$ comes 
from the fact that we round $b_i$ to the nearest power of two immediately
above prior to using Lemma~\ref{lemma:Ra96}) so that each key
is mapped to a distinct cell. This hash function
occupies $O(\log n)$ bits in total and is determined in $O(b_i^3)$ time. 
The hash function is determined as follows: for every bucket $i$ to 
which the set of mapped keys is $S_i$, we let $S'_i\subset [n^2]$ be the image 
of $S_i$ under the function $f_2$ (note that $f_2$ is injective on $S_i$). 
Whenever $b_i>\log n$ (we call such buckets large and the others small), 
use Lemma~\ref{lemma:Ra96} and $h_i$ will be the resulting hash function that maps 
every $x'$ in $S_i$ to a unique number in the range $[4(b_i)^2]$ . If $b_i\leq \log n$, we apply 
Lemma~\ref{lemma:Fusion_tree}, generating a hash function $h_{i,1}$ 
that injectively maps $S'_i$ to a set $S''_i\subset[2^{b_i-1}]$. 
We can easily simulate the instructions $\mathrm{MSB}$ and $\mathrm{PACK}$ 
using the four russians technique~\cite{ADKF75} after a preprocessing phase of $o(n)$ time. Since the arguments 
are of length $2\log n$ bits, we can use a table to simulate 
$\mathrm{MSB}$ on $4$ chunks of length $\lceil\log n/2\rceil$ and another table 
to simulate the $\mathrm{PACK}$ instruction on $8$ pairs of arguments
of length $\lceil\log n/4\rceil$. 
We then use Lemma~\ref{lemma:Ra96} on the set $S''_i$ to generate a hash function $h_{i,2}$. 
We let $h_i$ be such that $h_i(x)=h_{i,2}(h_{i,1}(x))$. 
Analysing the time to generate $h_i$ on a set $S'_i$, we can see that
the computation in Lemma~\ref{lemma:Ra96} takes $O(b^2_i\log n)\in O(b^3_i)$ if
$i$th bucket is large and $O(b^2_i\log (2^{b_i-1}))=O(b^3_i)$ otherwise, 
and the computation in Lemma~\ref{lemma:Fusion_tree} takes $O(b_i\log b_i)\in O(b^3_i)$. 
Each bucket stores a representation that consists 
in the number $b_i$ and the description of the hash function $h_i$ using $O(\log n)$ bits. 
We store an array $b[0..n-1]$ in which $b[i]=b_i$. We also store an array $B[0..n-1]$ such that $B[i]$ stores the description
of $h_i$. If the set $S_i$ is empty, then $B[i]$ will store a special value $\mathtt{null}$. 
The two arrays $b$ and $B$ together form the bucket representation. 
Note that the time to determine the bucket representation is $O(n)$ since 
$O(\sum_{i=0}^{n-1}(b_i)^3)=O(n)$~\footnote{We note 
that a hash function of the same form as that used in Lemma~\ref{lemma:Ra96} 
can be constructed directly in time $O((b_i)^2\log b)$ time~\cite{ruvzic2008uniform}, 
and thus avoiding the need to use the hash function of Lemma~\ref{lemma:Fusion_tree}. 
However the algorithm of~\cite{ruvzic2008uniform} and its analysis seem more 
complicated than those used in the lemmata.}. 

Finally the keys are stored in their associated cells as follows.
We use a table $\beta[0..N-1]$ of $N=\sum_{i=0}^{i=n-1} 4(b_i)^2=O(n)$ 
cells that store pairs $(x,c)$, 
where $x\in[u]$ is a position in $s$ and $c\in[\sigma]$ is a character. 
The cells are assigned as follows: 
For each position $x$ such that $s[x]\neq 0$ and $y=f_1(x)$, we store 
the pair $(x,s[x])$ in $\beta[h_y(x)+j]$,
where $j=\sum_{i=0}^{y-1} 4(b_i)^2$. 
For every position $k$ of $\beta$ for which no pair was assigned, 
we assign the special value $\mathtt{null}$ to $\beta[k]$. 
We call the resulting table $\beta$, the cell representation. 
Note that the resulting table has $O(n)$ cells each needing $\log\sigma+\log u$ bits. 

In the next phase player $A$ uses a systematic error correcting code 
on top of each of the arrays $b$, $B$ and $\beta$ capable
of correcting $k$ errors. The redundancies will be 
respectively $r(n,k,2^{\Theta(\log n)})$, $r(n,k,2^{\Theta(\log u)})$ and $r(n,k,2^{\Theta(\log u+\log\sigma)})$.

Finally player $A$ sends the description of the hash function $f$ and 
the redundancies built on top of the arrays $b$, $B$ and $\beta$
for a total of $\log u+\Theta(n,k,2^{\Theta(\log u+\log\sigma)})$ bits. 
Note that the time spent by player $A$ is randomized $O(n)$ in addition 
to the time needed to build the redundancies of the error correcting codes.

\subsection{Receiver protocol}

We now describe the recovery done by player $C$. Player $C$ starts by applying the hash function $f$ (which he got from player $A$) on the set $S'$ of keys consisting in positions of non-zeros in $t[0..u-1]$. He then assigns to each bucket $i$ the keys that are mapped to value $i$ by the hash function, and builds an array $b'[0..n-1]$ such that $b'[i]$ stores the number of keys assigned to the bucket number $i$. Then player $C$ uses the redundancy that player $A$ has built on top of $b$ to correct the array $b'$ into the array $b$. In the next step, for every bucket $i$ such that $b'[i]=b[i]$, player $C$ determines the hash function $h_i$ based on the keys assigned $S'_i$ to the bucket $i$ using exactly the same algorithm as that used by player $A$. He builds an array 
$B'[0..n-1]$ as follows. 
If $b'[i]\neq b[i]$, then $B'[i]$ will contain an arbitrary value. Otherwise, it will contain the description of the hash function $h_i$ (if $b[i]=0$ then $B'[i]$ will contain $\mathrm{null}$). Player $C$ can now recover the array $B$ from the array $B'$ using the redundancy that player $A$ has built on the array $B$. 
Next player $C$ builds a cell representation $\beta'$ based on the bucket representation of player $A$. That is, for every bucket $i$, apply the hash function $h_i$ on every key $x$ that is assigned to bucket $i$ and map it to position $\beta'[h_y(x)+j]$ where $j=2\sum_{i=0}^{y-1} (b[i])^2$. Note also that there could be collisions between the keys into each cell of player $C$. If more than one key is mapped to a cell then choose one of them arbitrarily and assign it to the cell. 
If a cell $i$ has no key assigned to it, then player $C$ sets $\beta'[i]=\mathtt{null}$. 
Then player $C$ can correct its own cell representation $\beta'$ using the redundancy 
that has been previously built by player $A$ on its own cell representation $\beta[0..N-1]$.

\subsection{Analysis}
We argue that the correction will work because each of the pairs of arrays $(b,b')$, $(B,B')$, ($\beta,\beta'$) will differ in at most $k$ positions. To see why, notice that the (at least) $n-k$ keys that are in the intersection of $S'$ and $S$ are mapped to the same buckets. That means that at most $k$ buckets of player $C$ could get a different set of keys from the corresponding buckets of player $A$. This implies that the correction will work correctly for the array $b'$. Moreover, the fact that the determination of the functions $h_i$ is fully deterministic implies that the same $h_i$ will be built for every bucket $i$ that has the same set of assigned keys for both players. Therefore the correction will also work out for the array $B'$. 
We argue that the cell representations $\beta$ and $\beta'$ also differ in at most $k$ positions. Given a bucket $i$, denote by $k_i$ the number of keys that differ between bucket number $i$ of player $A$ and bucket number $i$ of player $C$. Then at least $b_i-k_i$ keys will be assigned to the same cell for both players. Thus at most $k_i$ cells of bucket $i$ will differ and overall the content of at most $k=\sum_{i=0}^{i=n-1}(k_i)$ cells will differ between the two representations.

Note that the time spent by player $C$ is deterministic $O(n)$ in addition to the time needed to apply 
the error corrections. The analysis of the running time is trivial except for the step where player $C$ builds the array $B'[0..n-1]$. 
In that case, note that player $C$ searches a function $h_i$ only if $b'[i]=b[i]$, spending exactly the same time spent by  player $A$. Thus the total time player $C$ spends on building the array $B'$ is no larger than the time spent by player $A$ 
on building the array $B$.

\begin{theorem} 
\label{thm:det_set_large_univ_red}

We can solve the one-way set reconciliation problem   
with a protocol that transmits $O(\log u+r(n,k,2^{\Theta(\log\sigma+\log u)}))$
bits and that runs in randomized $O(n+t_e(n,k,2^{\Theta(\log\sigma+\log u)}))$
time for the sender side and deterministic time $O(n+t_d(n,k,2^{\Theta(\log\sigma+\log u)}))$
for the receiver side (the time becomes randomized if $t_d$ is randomized). 

\end{theorem} 

To get an efficient scheme we will use Reed-Solomon codes as our underlying error correcting code. 
We divide each character to be represented into $\Theta(\frac{\log u+\log\sigma}{\log n})$ chunks of $\log n$ bits each. Then we build $\Theta(\frac{\log u+\log\sigma}{\log n})$ 
different strings from the original string. 
The alphabet of the new strings will be of size $n$ and the character number $j$ of the string number $i$ will receive the chunk number $i$ of the character number $j$ in the original string. 

The redundancies will occupy a total of $\Theta(k(\log u+\log\sigma))$ bits of space. 

\begin{theorem} 
\label{thm:det_set_sol}
We can solve the one-way set reconciliation problem with 
a protocol that transmits $\Theta(k(\log u+\log\sigma))$ 
bits and succeeds with certainty. The
protocol runs in time 
$O(n\cdot \mathrm{polylog} (n)\cdot (\log u+\log\sigma))$ 
for both parties. 
The time is randomized for the sender and deterministic for the receiver. 
\end{theorem} 
\section{Improved reduction}
We will now present a better reduction (actually optimal) with reduced alphabet size. 
Since the reduction above is optimal whenever $u=n^{1+\Omega(1)}$ (i.e., the reduction 
implies a message size $O(k(\log u+\log\sigma))$ bits which is optimal up to constant factors),  
we will focus here only on the case $u\leq n^{3/2}$. 

\begin{theorem} 
\label{thm:det_set_small_univ_red}
We can solve the one-way set reconciliation problem  
when $n\geq u^{2/3}$ with a protocol that transmits $O(\log u+r(n,k,2^{\Theta(\log\sigma+\log(u/n))}))$
bits and that runs in randomized $O(n+t_e(n,k,2^{\Theta(\log\sigma+\log(u/n))}))$
time for the sender side and deterministic time $O(n+t_d(n,k,2^{\Theta(\log\sigma+\log(u/n))}))$
for the receiver side (the time becomes randomized if $t_d$ is randomized).

\end{theorem} 
The main ingredient we will use to reduce the term $\log u$ to $\log(u/n)$, 
is a technique known as quotienting~\cite{pagh2001low}. 
More in detail, we modify the previous scheme as follows. 
The first modification is to use only Lemma~\ref{lemma:3wise_hash}
to generate the function $f$ directly on the set $S$. Since 
$S\subset[u]\subseteq[n^{3/2}]$, we do not need to use 
Lemma~\ref{lemma:2wise_hash_no_prime} to first reduce the keys 
to universe $[n^2]$. 

The second modification is to use a different finite field 
in Lemma~\ref{lemma:3wise_hash}. 
Instead of using 
the finite field $\mathbb F_P$ for a prime $P\geq u$, 
we will use the finite field $\mathbb F_{q^2}$ for some 
prime $q=\Theta(\sqrt{u})$~\footnote{The main reason for the modification 
is that we will now need to solve equations over the finite field, 
which will require efficient support for inversions and square 
root operations. Those operations can be supported using lookup 
tables of $O(P\log P)$ and $O(q\log q)$ bits respectively on $F_P$
and $F_{q^2}$. Since $|u|$ can be as big as $n^{3/2}$, the lookup tables 
of $F_p$ will occupy $O(n^{3/2}\log n)$ bits while those 
for $F_{q^2}$ will occupy only $o(n)$ bits of space and can be computed in $o(n)$
additional time.}. 
More precisely the elements 
of $\mathbb F_{q^2}$ are polynomials of degree two over 
numbers from $[q]$ (and all operations on the numbers 
are computed modulo $q$) with all the operations 
on the polynomials computed modulo the polynomial $\gamma^2-A$, where $A$ 
is a quadratic non-residue for $\mathbb F_q$ (there is no $y$ 
such that $(y^2\bmod q)=A$). Note that a number $x\in [u]$ is now 
represented in $\mathbb F_{q^2}$ as the polynomial $x_1\gamma+x_0$, where $x_1=(x\;\mathrm{div}\;q)$
and $x_0=(x\;\bmod q)$. 

We redefine $f_2$ to be such that 
$f_2(x)=(f(x)\;\mathrm{div}\;n)$. 
This reduces the terms $\log u$ and $\log n$
to $\log(u/n)$ in the space used by the cell 
and bucket representations (this reduction, well-known 
in the literature is called \emph{quotienting}~\cite{pagh2001low}). That is array $b[0..n-1]$
will now use $\log(u/n)$ bits per entry since any bucket 
can get at most $u/n$ keys assigned to it. Now the set $S'_i$
is the image of $S_i$ under the function $f_2$, and the hash 
function $h_i$ occupies $O(\log (u/n))$ bits. 
For the cell representation, we will replace every key $x$
by pair $(I,f_2(x))$, where $I$ is a bit computed as follows. 
We solve the equations $aX^2+bX+c=f(x)$, which determines two 
solutions $X_0<X_1$. Then if $X_0=x$, we set $I=0$, otherwise, 
$X_1=x$ and we set $I=1$. Now, player $C$ can recover a key 
$x$ that is mapped to a bucket $i$ and a cell containing 
the pair $(I,j)$ by solving the equation $aX^2+bX+c=j\cdot n+i$, 
obtaining two numbers $X_0$ and $X_1$ and $x=X_I$. 
Solving the equation involves additions, subtractions, 
multiplication, inversion and square root operations. 
All the operations on $\mathbb F_{q^2}$ 
can be implemented using a constant 
number of operations from the same set of operations 
on the base field $\mathbb F_q$, even square root
and inversion~\cite{adj2012square}. 
All operations on $\mathbb F_q$ are trivially supported 
in constant time, except for square root and inversion which 
are implemented using lookup tables 
that use $O(n^{3/4}\log n)$ bits of space and are computed in time 
$O(n^{3/4})$. 
By combining theorems~\ref{thm:det_set_small_univ_red} and ~\ref{thm:det_set_large_univ_red},
we can state the following formal reduction:

\begin{theorem}
\label{thm:det_set_opt_red}
We can solve the one-way set reconciliation problem 
with a protocol that transmits $O(\log u+r(n,k,2^{\Theta(\log\sigma+\log(u/n))}))$
bits and that runs in randomized $O(n+t_e(n,k,2^{\Theta(\log\sigma+\log(u/n))}))$
time for the sender side and deterministic time $O(n+t_d(n,k,2^{\Theta(\log\sigma+\log(u/n))}))$
for the receiver side (the time becomes randomized if $t_d$ is randomized). 

\end{theorem}

The reduction is optimal (up to constant factors), 
since, on the one hand, even if player $A$ knows 
player $C$'s string, he needs to send the positions of the mismatches
and the mismatching characters using $\Omega(k(\log(u/k)+\log\sigma))$ bits, 
and on the other hand the protocol sends  
$\Theta(\log u)=\Theta(\log (u/k)+\log k)\leq \Theta(k\log (u/k))$ bits in addition
to the redundancies of the error correcting codes 
which can be $\Theta(k(\log (u/k)+\log\sigma))$ bits in the best case. 
The latter follows from the existence of error correcting codes that achieve 
optimal redundancy $\Theta(k(\log(n/k)+\log\sigma'))=\Theta(k(\log(n/k)+\log(u/n)+\log\sigma))$, 
in particular those attaining the 
Gilbert-Varshamov bound~\cite{gilbert1952comparison,varshamov1957estimate}.
Unfortunately, we do not know 
whether such codes admit $O(n\cdot \mathrm{polylog} (n))$ 
decoding time for all values of $k$ and $n$, 
and thus we can not use Theorem~\ref{thm:det_set_opt_red} 
to state a message-size optimal 
solution that would improve on Theorem~\ref{thm:det_set_sol}. 

\section{Concluding remarks}
It was implicit that the numbers $\log\sigma$, $\log u$ and $\log n$ are known to the algorithm. 
If that was not the case, then we need to have an implementation of the $\mathrm{MSB}$ operation. 
The operation can be simulated in constant time, using a constant number of standard 
instructions (including multiplications)~\cite{brodnik1993computation}.
Our reduction will also need to efficiently simulate divisions by $u$ and $n$. 
Division by $u$ is used in Lemma~\ref{lemma:2wise_hash_no_prime} and division 
by $n$ is used in Lemma~\ref{lemma:3wise_hash}. If both $u$ and $n$ 
are powers of two, then the division is just a right shift. 
Otherwise, the division (and modulo) can be simulated using the multiplication 
by the inverse. The latter can be precomputed, respectively, in $O(\log\log u)$ and 
$O(\log\log n)$ time (using the Newton-Raphson method). The term $O(\log\log n)$ is clearly within our time 
bound, but this is not necessarily the case for the term $O(\log\log u)$. 
We can avoid the problem by rounding $u$
to the nearest power of two immediately above (using the $\mathrm{MSB}$ operation). 
This will only increase the final size of the message by $O(k)$ bits. 

In the proof of Theorem~\ref{thm:det_set_small_univ_red}, we can replace the polynomial from Lemma~\ref{lemma:3wise_hash} by a 3-wise independent permutation, 
computed using the M\"{o}bius transform~\cite{alon2012almost} 
over the projective line $\mathbb F\cup \{\infty\}$, 
where $\mathbb F$ is any finite field. 
The projection is given by the formula $y=\frac{ax+b}{cx+d}$, where
$a,b,c,d$ are constants such that $ad-bc=1$. Its reverse can be computed through the formula 
$x=\frac{dy-b}{-cy+a}$. 
We can thus use $\mathbb F=\mathbb F_{q^2}$ and simulate the transform and its inverse 
in constant time (in particular simulating division in constant time over 
$\mathbb F_{q^2}$ by using operations on $\mathbb F_q$). Then in a cell, 
we will only need to store a value $f_2(x)$ instead of pair $(I,f_2(x))$, since 
we can now recover $x$ directly from $f(x)=f_2(x)\cdot n+f_1(x)$. 

Our modification of the two-level hashing scheme of~\cite{FredmanKS84}
may be of independent interest. 
It uses only $O(\log u)$ bits of randomness and only in the first level.
This is not the first result of the kind. In~\cite{dietzfelbinger1992polynomial}, 
a variant of the scheme of~\cite{FredmanKS84} is presented which uses only $O(\log n+\log\log u)$
bits of randomness.
We could use such a scheme as an alternative in our reduction. The sender would build
the structure, and then send the random bits that he uses. The main drawback of that approach 
is that the second level is randomized too, in particular, the construction time
of the functions of the buckets depends on the keys mapped to the bucket (and not 
only their number as in our case). 
As a consequence, there is no guarantee that the receiver will spend a total 
deterministic linear time for building the second level (the time is only expected linear). 
We could also have slightly simplified our scheme by only implementing 
set reconciliation with alphabet $\{0,1\}$ and simulating any alphabet
~.\textsuperscript{\ref{big_to_binary_alpha}}. The drawback of doing so 
is an increase in the message sice, since now the $\log u$ term 
is replaced by $\log u+\log\sigma$ in the description of the global hash function 
and in the redundancy associated with array $B$. 
A future research direction is to reduce the constants associated with 
our scheme and to make the reduction run in deterministic time for the 
sender side.  

\section*{Acknowledgements}
This work was initiated when the author was visiting MADALGO 
(Aarhus University) in Spring 2012. The author wishes to thank Hossein Jowhari 
and Qin Zhang for the initial fruitful discussions that motivated 
the problem and the anonymous referees whose suggestions greatly improved 
the presentation of the paper. 
\bibliographystyle{plain}
\bibliography{set_reconciliation.bib}

\end{document}